\documentclass[draft]{agujournal2019}
\usepackage{url} %this package should fix any errors with URLs in refs.
\usepackage{lineno}
\usepackage[finalnew]{trackchanges} %for better track changes. finalnew option will compile document with changes incorporated. , inline
\usepackage{soul}

\usepackage{amsmath}
\usepackage{amssymb}

\draftfalse

\journalname{Geophysical Research Letters}

\begin{document}

%% ------------------------------------------------------------------------ %%
%  Title
%
% (A title should be specific, informative, and brief. Use
% abbreviations only if they are defined in the abstract. Titles that
% start with general keywords then specific terms are optimized in
% searches)
%
%% ------------------------------------------------------------------------ %%

% Example: \title{This is a test title}

\title{The fate of liquids trapped during the Earth's inner core growth}

%% ------------------------------------------------------------------------ %%
%
%  AUTHORS AND AFFILIATIONS
%
%% ------------------------------------------------------------------------ %%

% Authors are individuals who have significantly contributed to the
% research and preparation of the article. Group authors are allowed, if
% each author in the group is separately identified in an appendix.)

% List authors by first name or initial followed by last name and
% separated by commas. Use \affil{} to number affiliations, and
% \thanks{} for author notes.
% Additional author notes should be indicated with \thanks{} (for
% example, for current addresses).

% Example: \authors{A. B. Author\affil{1}\thanks{Current address, Antartica}, B. C. Author\affil{2,3}, and D. E.
% Author\affil{3,4}\thanks{Also funded by Monsanto.}}

\authors{M. Lasbleis\affil{1,2}, M. Kervazo\affil{1}, G. Choblet\affil{1}}

% \affiliation{1}{First Affiliation}
% \affiliation{2}{Second Affiliation}
% \affiliation{3}{Third Affiliation}
% \affiliation{4}{Fourth Affiliation}

\affiliation{1}{Laboratoire de
Plan\'etologie et G\'eodynamique, LPG, UMR 6112, CNRS, Universit\'e de Nantes, Universit\'e d'Angers, France}
\affiliation{2}{Earth-Life Science Institute, Tokyo Institute of Technology, Meguro, Tokyo 152-8551, Japan}

%(repeat as many times as is necessary)

%% Corresponding Author:
% Corresponding author mailing address and e-mail address:

% (include name and email addresses of the corresponding author.  More
% than one corresponding author is allowed in this LaTeX file and for
% publication; but only one corresponding author is allowed in our
% editorial system.)

% Example: \correspondingauthor{First and Last Name}{email@address.edu}

\correspondingauthor{Marine Lasbleis}{marine.lasbleis@univ-nantes.fr}

%% Keypoints, final entry on title page.

%  List up to three key points (at least one is required)
%  Key Points summarize the main points and conclusions of the article
%  Each must be 100 characters or less with no special characters or punctuation and must be complete sentences

% Example:
% \begin{keypoints}
% \item	List up to three key points (at least one is required)
% \item	Key Points summarize the main points and conclusions of the article
% \item	Each must be 100 characters or less with no special characters or punctuation and must be complete sentences
% \end{keypoints}

\begin{keypoints}
\item We calculate the residual melt fraction of liquid in the inner core for different growth scenarios
\item An uppermost mushy layer is maintained, of thickness of about 1\,km for a viscosity of $10^{20}$\,Pa.s
\item Supercooling at the center of the core cannot have been larger than 100\,K
\end{keypoints}

%% ------------------------------------------------------------------------ %%
%
%  ABSTRACT and PLAIN LANGUAGE SUMMARY
%
% A good Abstract will begin with a short description of the problem
% being addressed, briefly describe the new data or analyses, then
% briefly states the main conclusion(s) and how they are supported and
% uncertainties.

% The Plain Language Summary should be written for a broad audience,
% including journalists and the science-interested public, that will not have 
% a background in your field.
%
% A Plain Language Summary is required in GRL, JGR: Planets, JGR: Biogeosciences,
% JGR: Oceans, G-Cubed, Reviews of Geophysics, and JAMES.
% see http://sharingscience.agu.org/creating-plain-language-summary/)
%
%% ------------------------------------------------------------------------ %%

%% \begin{abstract} starts the second page

\begin{abstract}
The growth history of the inner core is inherently linked to the thermal history of the Earth. The crystallization of the inner core may have been delayed by supercooling, and went through an initial phase of fast growth after the nucleation barrier has been passed, but so far no evidence or constraint has been proposed to time this possible event. 
With two-phase flow dynamics, we explore the effect of different growth scenarios for the inner core to determine their effects on its liquid fraction structure. Seismic observations on the melt fraction inside the inner core at present limit the porosity to a maximum of 10\% of liquid. 
Our model provides constraints for the delay in nucleation compared to the time where the first crystal may have started to nucleate, and we find that the supercooling cannot have exceeded 100\,K. 
\end{abstract}

\section*{Plain Language Summary}
The magnetic field of the Earth is maintained by the turbulent convection in the liquid outer core. As the inner core grows, latent heat and light elements are released into the liquid outer core, generating upwelling of material and driving the convection. The timing for the nucleation of the inner core, starting the dynamo as we know it today, has been challenging to assess from the paleomagnetic record. Recent work proposed that the nucleation of the first crystal may have been delayed by a supercooling effect, as the core is lacking adequate solid particles as a substrate for crystallization. Here we propose to test the effect of such  late nucleation on the melt fraction trapped inside the inner core. In the supercooling hypothesis, the growth of the inner core presents two stages, starting with a very fast growth. We show that such a scenario would prevent melt to flow out of the inner core. Since seismic observations do not show the presence of large amounts of liquid {with} depth, we argue that this is the first evidence to discard the possibility of a supercooling larger than 100\,K.

%% ------------------------------------------------------------------------ %%
%
%  TEXT
%
%% ------------------------------------------------------------------------ %%

%%% Suggested section heads:
% \section{Introduction}
%
% The main text should start with an introduction. Except for short
% manuscripts (such as comments and replies), the text should be divided
% into sections, each with its own heading.

% Headings should be sentence fragments and do not begin with a
% lowercase letter or number. Examples of good headings are:

% \section{Materials and Methods}
% Here is text on Materials and Methods.
%
% \subsection{A descriptive heading about methods}
% More about Methods.
%
% \section{Data} (Or section title might be a descriptive heading about data)
%
% \section{Results} (Or section title might be a descriptive heading about the
% results)
%
% \section{Conclusions}

\section{Introduction}

The growth history of the inner core is inherently linked to the thermal history of the Earth's core and mantle. Understanding it would help unravel the links between the magnetic history of the Earth and the thermal and compositional structure of the core. The evidence for the timing of the inner core nucleation and growth in the paleomagnetic records is sparse \cite{biggin2015palaeomagnetic, smirnov2016palaeointensity, bono2019young} and the changes in the magnetic field may have been moderate \cite{landeau2017signature}, with potential for a change in the field geometry \cite{Driscoll2016Simulating}.
As the Earth cools down, the partial crystallization from the liquid outer core releases latent heat and light elements at the bottom of the outer core, thus providing the major source of energy for vigorous convection \cite{verhoogen1961heat, braginsky1963structure, gubbins1977energetics, lister1995strength}. 
Considering that the inner core boundary (ICB) is anchored at the intersection between the outer core adiabatic temperature and the melting temperature of the iron alloy, the age of the inner core has been estimated from 1.5\,Gyr to 200\,Myr \cite <e.g.>[]{buffett1996thermal, Labrosse2001TheCore, buffett2003thermal,  nimmo2007energetics, monnereau2010lopsided, Labrosse2015ThermalConductivity}. However, the recent work of \citeA{Huguet2018EarthsParadox} challenged this hypothesis. \citeA{Huguet2018EarthsParadox} argue that the nucleation of the inner core requires to overcome a crystallization barrier for supercooling \cite{shimizu2005crystallization, davies2019assessing}, and that it is likely that the formation of the first crystals was delayed. The subsequent initial growth stage was then much faster than in the classical scenario in order to reach the same radius at present. So far, no evidence for such delayed nucleation, which timing might also be linked to a catastrophic event in the Earth's mantle, have been proposed.

Seismic and mineral physics studies have shown that the core is made of a mixture of iron and nickel and some light elements \cite{birch1961composition, birch1964density, mcdonough1995composition}, which exact composition is still unknown \cite{poirier1994light, vovcadlo2007ab}. 
The seismic structure of the solid inner core is complex and difficult to access, exhibiting heterogeneities in seismic properties {at global and regional scales}, both radially and horizontally \cite <see for review:>{deuss2014heterogeneity}.  The existence of an innermost inner core has been proposed by several studies, with the observation of a different cylindrical anisotropy relative to the outer part, but the size of this zone and the associated anisotropy orientation are still debated \cite{ishii2002innermost, wang2015equatorial,romanowicz2016seismic, wang2018support}.
According to the PREM model \cite{dziewonski1981preliminary}, the inner core is associated to low shear wave velocity and high attenuation. These features have been shown by the recent detection of PKJKP waves \cite{tkalvcic2018shear} to be even more pronounced, revealing a soft inner core.
In the bulk of the inner core, \citeA{singh2000presence} showed that a volume fraction of liquid of 3-10\% may explain the low shear wave velocityalthough the presence of liquid is not necessarily required \cite{tkalvcic2018shear}. 
Below the ICB, the existence of lateral variations of seismic properties may also show the presence of patches of mushy zones \cite{tian2017seismological}, while the ICB itself is globally sharp, with a thickness of at least less than  3\,km \cite{koper2003constraints}. 

The co-existence of liquid and solid iron inside the inner core has been proposed early on also from theoretical studies. 
\citeA{loper1981study} and \citeA{fearn1981structure} suggested that either a slurry zone or a mushy zone could exist due to crystallization process at the ICB, which thickness may extend to the very center of the Earth. Liquid may be trapped near the surface of the inner core due to a morphological instability of the solidification front, leading to dendritic crystals \cite <e.g.>[]{mullins1963morphological, davis2001theory, deguen2007existence} or by sedimentation of crystals \cite{Sumita1996AGrowth}. The depletion in light elements of the remaining liquid near the ICB will prevent its solidification, as  the eutectic temperatures of most iron alloys at inner core pressure are thought to be well below the temperature at the center of the inner core \cite{morard2014properties}. The liquid is then expelled from the solid matrix as it compacts under its own weight \cite{Sumita1996AGrowth} or due to convective instabilities. 

Compaction of a solid matrix has been extensively studied in the geosciences literature for processes such as magma chamber solidification \cite{McKenzi1984TheRock, mckenzie2011compaction}, magma extractions \cite <e.g.>[]{rabinowicz2004melt}, planetary evolution from accretion \cite <e.g.>[]{scheinberg2016core, hier2017origin} and petroleum transfer in sediments \cite <e.g.>[]{appold2002numerical}, but also for the evolution of the inner core \cite{Sumita1996AGrowth}. In the context of the inner core growth, a notable difference from most compaction studies comes from the linear dependence of gravitational acceleration on radius in the core, due to the self-gravitation.
\citeA{Sumita1996AGrowth} have shown that, in simple growth histories, the trapping of liquid in the inner core can be estimated by comparing the rate of compaction and the rate of solidification. If the compaction is faster than the sedimentation at the ICB, then compaction will be efficient, and little fluid will remain trapped. \citeA{Sumita1996AGrowth} have proposed that the Earth's inner core falls into this category, thus preventing significant residual melt in its bulk. 
In the framework of supercooling, an initial stage of very fast growth would modify drastically the amount of liquid trapped in depth. 

We propose here that this may be a key to decipher the existence and the timing of an episode of supercooling.
In this paper, we revisit the model of \citeA{Sumita1996AGrowth} for the evolution of a compacting inner core and study the effect of an episode of late crystallization on the porosity structure of the inner core. The comparison with seismic observations allows us to propose constraints on the growth history of the inner core, and discard the possibility of a late episode of crystallization.

\section{Model for compaction and growth of an inner core}
\subsection{Constitutive equations}

Following  \citeA{Sumita1996AGrowth}, we model the compaction of the inner core as a two-phase flow system where the dynamics are driven solely by the density difference between liquid and solid. Thermal and compositional contributions to buoyancy are not considered, discarding convective instabilities and allowing for a 1-D processing. Both phases are treated as viscous fluids, with a liquid viscosity several orders of magnitude smaller than the solid one. The set-up in this study is a self-gravitating two-phase sphere that grows from sedimentation of solid material at its surface.  

Similarly to recent works on two-phase flow dynamics \cite{Bercovici2001ATheory,Ricard2001ATension, Sramek2007ModeleDifferenciation}, we consider equations for the liquid and solid phases, assuming that both phases are incompressible, with constant viscosity and density. 
The subscripts $f$ and $m$ refer to liquid and solid matrix phases respectively, and $\phi$ is the volume fraction of liquid. The volume average of a quantity $A$ over a parcel of two-phase mixture is noted as $\bar{A} = \phi A_f+(1-\phi) A_m$ and the difference of a quantity $A$  between the solid and the liquid phase as $\Delta A = A_m-A_f$. We note $v_i$ the velocity of the phase $i$, $\mu_i$ the viscosity, $\rho_i$ the density and $P_i$ the pressure. 

The conservation equation of mass for the solid and liquid phases are combined to obtain an equation for the solid and liquid velocities
\begin{equation}
 \boldsymbol{\nabla} \cdot [(1-\phi) \mathbf{v}_m+\phi \mathbf{v}_f]=0.
\label{eq:solid+liquid}
\end{equation}

The momentum conservation equations for the liquid and the solid matrix are written as a generalized Darcy law and an equation for the matrix deformation, where $\mu_f\ll\mu_m$  \cite{Bercovici2001ATheory, Sramek2007ModeleDifferenciation} 
\begin{equation}\label{eq:Darcy}
\mathbf{0} = -\phi \boldsymbol{\nabla} P_f -\rho_f \phi \mathbf{g}+\frac{\phi^2 \mu_f}{k(\phi)}\Delta \mathbf{v},
\end{equation}
\begin{equation}\label{eq:matrix}
-\boldsymbol{\nabla} [(1-\phi) \Delta P] +(1-\phi) \Delta \rho \mathbf{g} +\boldsymbol{\nabla} \cdot [(1-\phi)\boldsymbol{\tau}_m] -\frac{\phi^2 \mu_f}{k(\phi)}\Delta \mathbf{v} +\Delta P \boldsymbol{\nabla} \phi = \mathbf{0}, 
\end{equation}
where the permeability chosen as $k(\phi) = k_0 \phi^n$, with $k_0$ a permeability coefficient and n an exponent chosen between 2 and 3; $g$ is the acceleration of gravity, assumed to be linear in radius and constant with time; $\boldsymbol{\tau}_m=\mu_m (\boldsymbol{\nabla}\mathbf{v}_m+[\boldsymbol{\nabla}\mathbf{v}_m]^T-2/3\boldsymbol{\nabla}\cdot \mathbf{v}_m\mathbf{I})$ is the viscous deviatoric stress tensor of the solid matrix. In this work, we will consider $n=3$ to be consistent with \citeA{Sumita1996AGrowth}. Assuming  a micro-mechanical model as $\phi \Delta P = -K \mu_m \boldsymbol{\nabla}\cdot \mathbf{v}_m$ with K a geometric factor of order one \cite{Sramek2007ModeleDifferenciation}, we combine equations \eqref{eq:Darcy} and \eqref{eq:matrix} as 
\begin{equation}
\frac{\phi \mu_f}{k(\phi)}\Delta \mathbf{v} = \boldsymbol{\nabla} \cdot \left ( (1-\phi){\boldsymbol{\tau}_m}\right )+ \boldsymbol{\nabla} \cdot \left ( \frac{1-\phi}{\phi}K\mu_m \boldsymbol{\nabla} \cdot (\phi \Delta \mathbf{v}) \right ) -(1-\phi)\Delta \rho \mathbf{g}.
\label{eq:moment_dim}
\end{equation}

Adding the conservation of mass of the solid phase, where we neglect any phase change and thus the exchange of mass due to melting or freezing, 
\begin{equation}
    \frac{\partial (1-\phi)}{\partial t}+\boldsymbol{\nabla} \cdot [(1-\phi) \mathbf{v}_m]=0, 
    \label{eq:cons_mass_solid}
\end{equation}
the set of equations \eqref{eq:solid+liquid}, \eqref{eq:moment_dim} and \eqref{eq:cons_mass_solid} provides a complete set of equations to solve for the porosity $\phi$ and the velocities of the solid and fluid phases.

\subsection{1D dimensionless equations}

We introduce characteristic scales for non-dimensionalization, choosing the compaction length scale $\delta = \sqrt{k_0\mu_m/\mu_f}$ \cite{McKenzi1984TheRock} as the characteristic length scale. The velocities are scaled with the Darcy velocity $V_D= | \Delta \rho | g_0 k_0/\mu_f $ where $|\Delta \rho| $ is the absolute value of the density difference between liquid and solid and $g_0$ the acceleration of gravity at the inner core boundary. Time is scaled with $\delta/V_D$. 

 The inner core is modeled as a sphere where no horizontal flows are allowed and all space dependencies of variables are radial dependencies. Thus, equation \eqref{eq:solid+liquid} gives $v_m = \phi \Delta v$. This allows us to write two equations for the matrix velocity $v_m(r)$ and the porosity $\phi(r)$ from the momentum equation \eqref{eq:moment_dim}  and mass conservation \eqref{eq:cons_mass_solid}  in non-dimensionalized form as 
\begin{gather}
\frac{\partial (1-\phi)}{\partial t} + \frac{1}{r^2} \frac{\partial}{\partial r} (r^2 (1-\phi) v_m)= 0, 
\label{eq:masse_solide_1d}\\
  \frac{\partial}{\partial r} \Bigg (\frac{(K+\frac{4}{3} \phi) (1-\phi)}{\phi} \frac{1}{r^2}  \frac{\partial}{\partial r} (r^2 v_m) \Bigg) + 4 \frac{v_m}{r} \frac{\partial \phi}{\partial r} + (1-\phi) s  \frac{r}{{R}_\text{ic}(\tau_\text{ic})} - \frac{v_m}{ \phi^n} = 0,
  \label{eq:moment_1d}
\end{gather}where the variables $t$, $r$,  $v_m$ are (from now) dimensionless, $s=\Delta \rho / | \Delta \rho |$ is the sign of the density difference and $R_\text{ic}(\tau_\text{ic})$ is the dimensionless radius of the inner core at present time $\tau_{ic}$. For the calculations, we choose $K=1$. 

\subsection{Boundary conditions and growth of the inner core}

The equations are solved between $r=0$ (referred as the center) and $r=R_{ic}(t)$ (referred at the ICB), with boundary conditions allowing for vertical flows to cross the ICB. The mechanical boundary conditions at the ICB are 
\begin{align}
v_m(r=R_\text{ic}(t)) +v_s  = \dot{R}_\text{ic}(t) , \label{eq:BC1}\\
\frac{\partial v_m}{\partial r}(r=R_\text{ic}(t)) = 0, \label{eq:BC2}
\end{align}
with $\dot{R}_\text{ic} (t)$ the {growth rate of the inner core}.

The porosity is set to $\phi_0$ at the ICB and $0$ at the center. $\phi_0$ is the porosity built when new material is added to the inner core by its growth. Its value should be similar to the rheological transition (see for example \citeA{renner2000rheologically}) from a slurry-type behavior with particles free in the liquid to a mush-type behavior with fully connected matrix and fluid, however, such a value is difficult to assess. Considering a sedimentation process, the values for random close packing of spheres are given from 0.25 to 0.47 volume fraction of liquid, and we will consider a value similar to \citeA{Sumita1996AGrowth}, with $\phi_0 = 0.4$. The rheological transition in the case of a dendritic-type mush at the ICB may be obtained for higher porosity values due to the large size of dendrites \cite{bergman1994chimneys, deguen2007existence}.

\begin{figure}[htp!]
\includegraphics[width=0.5\textwidth]{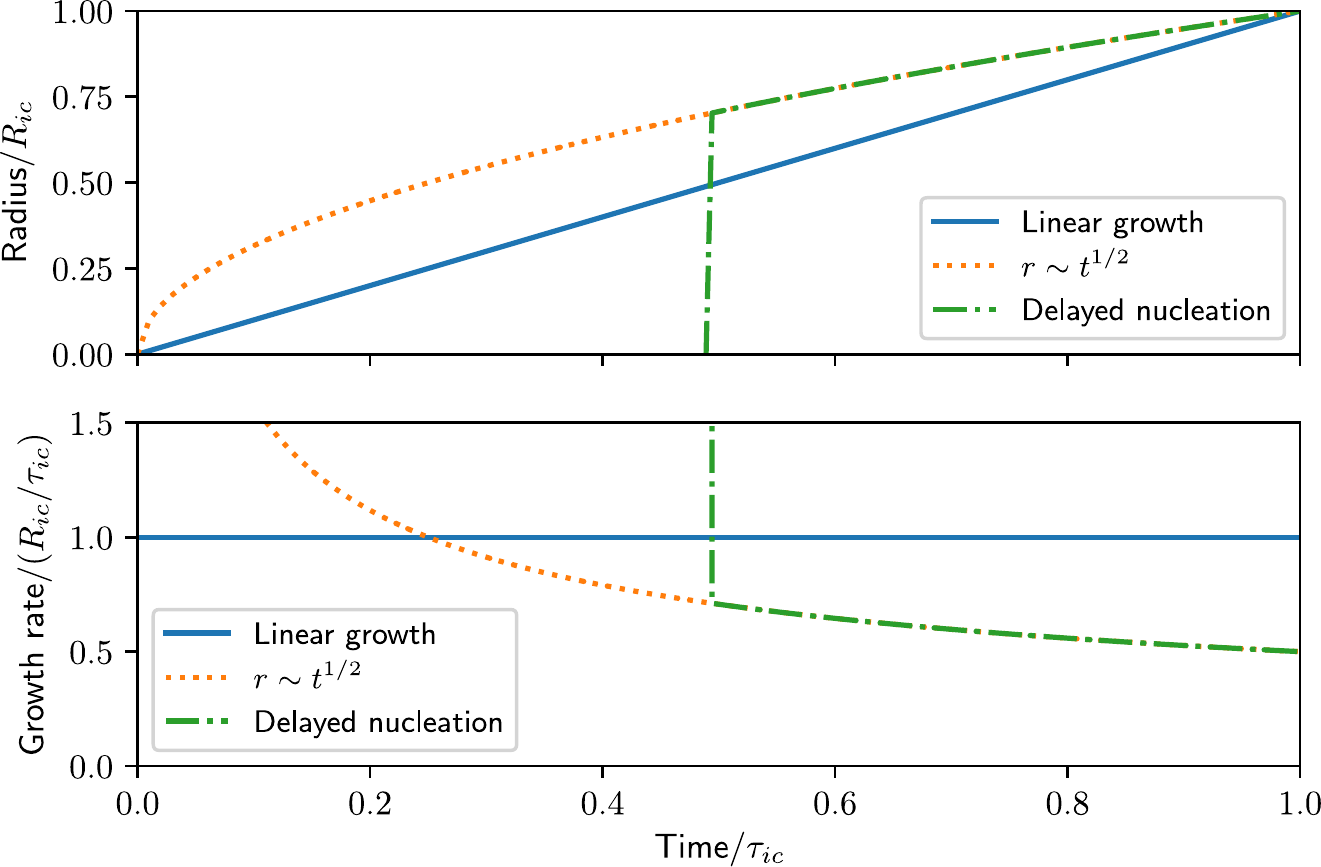}
\caption{Time evolution of the radius {(top)} and growth rate {(bottom)} in the different growth scenarios considered in this work: linear (solid blue), $R_\text{ic}(t)\sim \sqrt{t}$ (dotted orange) and two-stage growth with delayed nucleation (dashed green) \cite{Huguet2018EarthsParadox}. $\tau_\text{ic}$ is here the age of the inner core in the case of $R_\text{ic}(t)\sim t^{1/2}$ growth.}
\label{fig:growth}
\end{figure}

The time evolution of the radius of the inner core $\dot{R}_{ic}(t)$  is fixed by the growth scenario and  the sedimentation rate $v_s$ is given by $v_s(t) = \dot{R}_\text{ic}(t)-v_m (r=R_\text{ic}(t))$. Several growth scenarios for the inner core (see Figure \ref{fig:growth}) are implemented: (1) linear growth is first considered to derive a range of possible dynamical regimes, (2) $R_{ic}(t) =  R_\text{ic}(\tau_\text{ic}) \sqrt{t/\tau_\text{ic}}$  based on classical models of the thermal evolution of the core  \cite{buffett2003thermal,Labrosse2003ThermalCore} and (3) a very sudden growth followed by the classical evolution in order to consider the supercooling effect on the inner core growth \cite{Huguet2018EarthsParadox}. In the supercooling scenario, the crystallization does not start when the temperature of the core reaches the freezing temperature at the center, but after the temperature is low enough to overcome the crystallization barrier. We simplify the evolution of the radius of the inner core during a growth modified by supercooling, describing it with two stages: a first stage of super-fast growth with constant growth rate to reach the theoretical size of the inner core that would follow a simple $R_\text{ic}(t)\sim \sqrt{t-t_0}$ and then a second stage similar to the canonical scenario (see Figure \ref{fig:growth}). $t_0$ is defined as the time delay compared to the canonical model (2), so that the end time of the calculation is still $\tau_\text{ic}$. As discussed in \citeA{Huguet2018EarthsParadox}, the nucleation of the inner core and thus the start time $t_0$ of the fast growth period is controlled by an external catastrophic event able to overtake the nucleation barrier. The time delay between the theoretical beginning of the growth and the actual beginning is thus an unknown variable in this problem.

\subsection{Numerical model}

The equations \eqref{eq:masse_solide_1d} and \eqref{eq:moment_1d} are solved numerically using finite differences on a regular grid for the different growth scenarios and for various values of the compaction length{ by varying the total size of the system $R_{ic}$}. The porosity is advected by equation \eqref{eq:masse_solide_1d} using a flux limiter scheme, allowing for low numerical dissipation and non-oscillatory behavior \cite{Sramek2010AFormation}, with the velocity $v_m$ being calculated from \eqref{eq:moment_1d}. The growth of the inner core is obtained by adding grid points when required, with radius of the inner core fixed by the chosen growth scenarios as presented in Figure \ref{fig:growth}. The grid is initialized with five grid points, which values are set to the initial porosity.

\section{Results}

We present here the results of our set of runs. We explored a large range of parameters to cover the ones relevant to the Earth's inner core and further, to describe the different dynamical regimes. For each run, we calculate the time evolution of the porosity and extract two important properties of the profiles: the volume average of the porosity and the thickness of the uppermost layer, which is still enriched in liquid. {In some cases, the liquid accumulates in small volumes, creating waves similar to solitary waves propagating upwards, which dynamics have been extensively studied {\cite <e.g.>[]{scott1984magma, richard2012solitary}}.}

\subsection{Compaction without growth}

We first consider the simple case of an inner core reaching its final radius without any compaction occurring, and turning on the compaction after. The top boundary conditions allow for material to cross the ICB, decreasing the average liquid porosity as the matrix compacts. In this system, we vary the ratio of the compaction length $\delta$ and the radius of the inner core $R_\text{ic}$ and let the inner core compacts under gravity. Defining the compaction time as the time required to reach an average porosity half of the initial porosity (from 0.4 to 0.2), we obtain that it scales with $R_\text{ic}/V_D$ for a compaction length larger than $2 R_\text{ic}$ and  with $ \delta^2 /(R_\text{ic} V_D)$ for a compaction length lower than $0.2R_\text{ic}$. A minimum compaction time is obtained for a compaction length of about $0.6R_\text{ic}$. For small compaction lengths, the efficiency of compaction is weakened by the propagation of solitary waves with length-scales similar to the compaction length. For large compaction lengths, the compaction is inefficient as $\delta$ is larger than the typical size of the system. 
%For $\delta << 2 R_\text{ic}$, the compaction efficiency does not depend strongly on $\delta$. 

\subsection{Constant growth rate}

\begin{figure}[htp!]
    \centering
    \includegraphics[width=\textwidth]{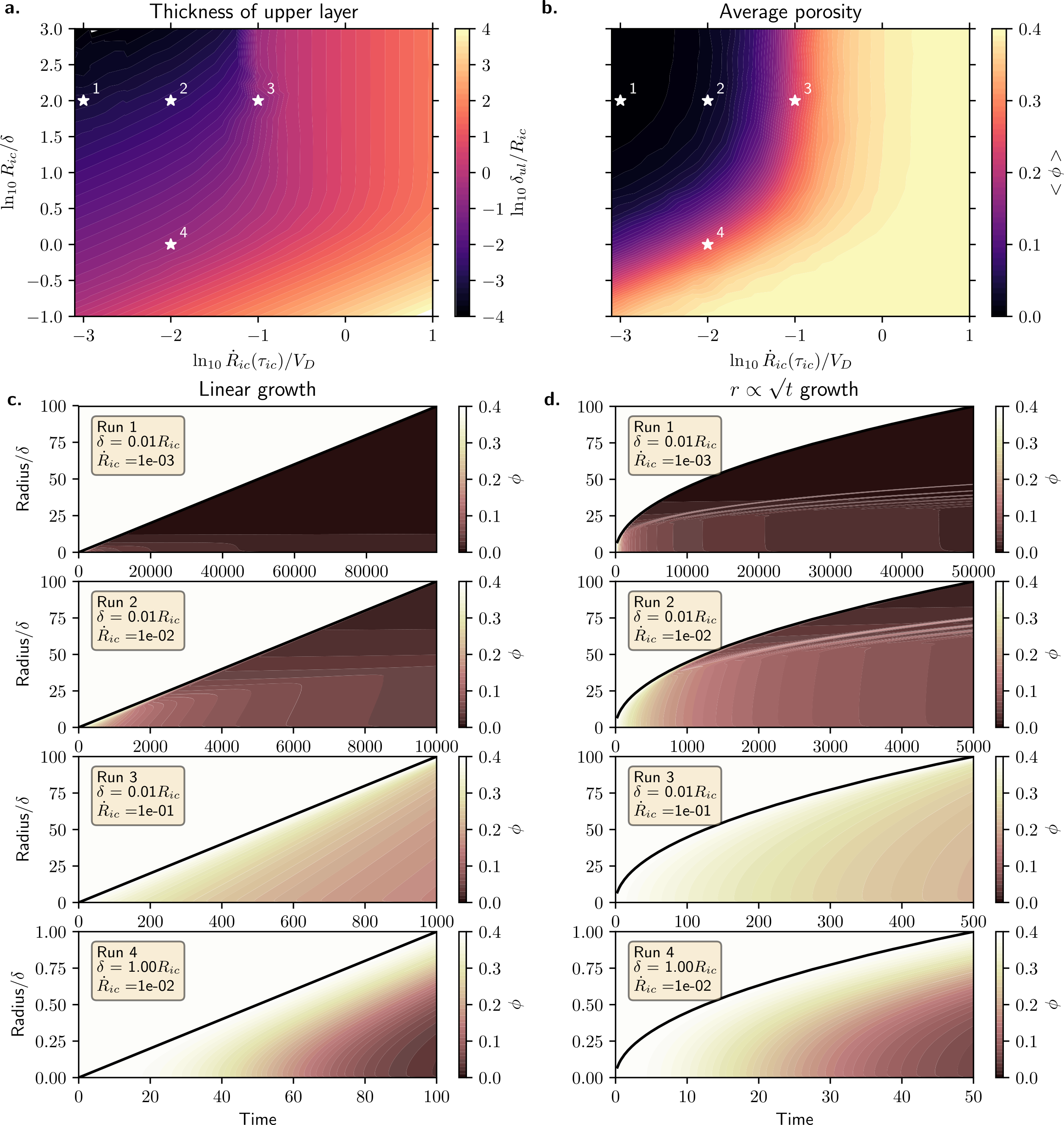}
    \caption{[a] Thickness of the uppermost mushy layer and [b] volume-average porosity for a linear growth as function of the final growth rate and inner core radius. [c] Time evolution of the porosity for the different runs highlighted by the white stars symbols in the panels [a] and [b]. 
    [d] Time evolution of the porosity for classical growth scenarios $R_\text{ic}(t)\sim \sqrt{t}$, for the same parameters as highlighted by the white stars in [a] and [b]. The growth rate is chosen at the end of the inner core growth $\tau_\text{ic}$.
    }
    \label{fig:compaction}
\end{figure}

Exploring the parameter space for constant growth rates gives a good idea of the different regimes, which depends on the values of the growth rate and the compaction length. 
Figure \ref{fig:compaction}[a-b] presents a summary of the results for linear growth rate. It shows the final thickness of the uppermost layer and the average porosity for a large range of compaction lengths and growth rates.  
%the result of the calculation of the thickness of the upper layer and the average porosity when the radius of the sphere reaches the final radius.
%Calculating the thickness of the upper layer and the average porosity when the radius of the sphere reaches the final radius, we obtain the Figure \ref{fig:compaction} (top). 
Two main regimes are observed: a compacted regime for low growth rates and small compaction length, and a non-compacted regime where the growth rate is larger than the Darcy velocity or where the compaction length is larger than the size of the system. The time evolution of the porosity profile for some parameters is also highlighted in Figure \ref{fig:compaction}[c]. 

For the Earth's inner core, the region of interest in the phase diagram is the one corresponding to a growth rate slower than the Darcy velocity, as it is where the porosity is significantly smaller than the initial porosity. We find that in that case, the porosity profile presents a compacted center and a non-compacted layer at the top, referred {to} as an uppermost mushy layer. The top region presents a profile similar to a half solitary wave, propagating at the same velocity than the growth rate, while some solitary waves can be observed for the smallest growth rate, propagating upwards but slower than the growth rate and originating from the center. 

The thickness of the upper layer $\delta_{ul}$\add{, which is the layer enriched in liquid, } is the typical thickness of  a solitary wave propagating in the system. It can be estimated for a Cartesian-geometry approximation similarly to the calculation in \citeA{Sumita1996AGrowth}. In our set of equations, we find that the thickness of the uppermost layer does not depend on the parameter $n$ and $\delta_\text{ul} \sim \delta \left ( \dot{R}\right ) ^{0.5}$, which is confirmed by our calculations\add{ (see Figure S2)}. From our calculations, the scaling law for the average porosity is $<\phi>\sim \dot{R}_\text{ic}$, for a compaction length tending towards 0, which is similar to the base porosity of the solitary wave profile obtained by \citeA{Sumita1996AGrowth}. 

{A difference in the initial porosity $\phi_0$ do not change the slopes of the scaling law shown here. For $\phi_0$ larger than 0.4, the scaling law for the thickness of the upper most layer is similar, and the porosities are obtained up to 50\% lower for initial porosity as high as 0.8.}

\subsection{Classical inner core growth}

To estimate the compaction in a more realistic set-up, we consider a growth history of the inner core as $R_\text{ic}(t)= R_\text{ic}(\tau_\text{ic})\sqrt{t/\tau_\text{ic}}$, where the growth rate is $\sim t^{-1/2}$ \cite{buffett2000earth, Labrosse2001TheCore}.  In this case, the growth rate is fast at first, and slowly decreases, as seen in Fig. \ref{fig:growth}. The porosity evolution is shown {in} Figure \ref{fig:compaction}[d], for the cases corresponding to the same parameters than shown by the white stars on the regime diagram {in}  Figure \ref{fig:compaction}[a-b]. 

For small compaction lengths, the system evolves thus from the absence of compaction due to a too-fast growth towards an efficient compaction. {As seen on Figure {\ref{fig:compaction}}[d], a}t the transition time, small packets of solitary waves start to propagate upwards, separating two regions: an innermost region where the scaling law for the porosity is $\phi\sim \dot{R}^{0.5}_\text{ic}$ and an outermost region of slightly lower porosity where the scaling law is $\phi\sim \dot{R}_\text{ic}$. In the upper region, the base porosity is set by the solitary wave attached to the top of the inner core, as additional material is advected upwards by solitary waves,  while the bottom region is at Darcy balance. \citeA{Sumita1996AGrowth} discussed the effect of the Darcy balance in a gravity field proportional to the radius. In that case, the balance is achieved for a constant porosity, which decreases with time. From the mass conservation equations, we obtain an evolution for the porosity at the center for any value of $n$ as $\partial \phi / \partial t \sim \phi^n/R_\text{ic}(\tau_\text{ic})$. The residual porosity at the center $\phi_c$ scales thus as $\phi_c\sim \sqrt[1-n]{t}/R_\text{ic}(\tau_\text{ic})$, confirmed by the results of the numerical models. For the case of linear growth, these two regions could not co-exist in a single run, as the growth rate is constant. 

We obtain finally that the scaling law for the thickness of the upper layer is identical to the linear case, as the uppermost layer structure is controlled by the recent growth properties. The average porosity does not follow a simple scaling law, as it is a combination of the two regions. For large compaction lengths ($\delta \gtrsim R_{ic}$), the average porosity tends towards the residual porosity at the center $\phi \sim \dot{R}^{0.5}_\text{ic}$, while at low compaction length ($\delta \lesssim R_{ic}$) it tends towards the porosity at the base of the upper layer $\phi \sim \dot{R}_\text{ic}$. Effectively, small compaction lengths and slow growth present efficient compaction, with an average porosity being less than 10\% for growth rates lower than $0.03\, V_D$,

\subsection{Growth after a supercooled event}

\begin{figure}[htp!]
    \centering
    \includegraphics[width=\textwidth]{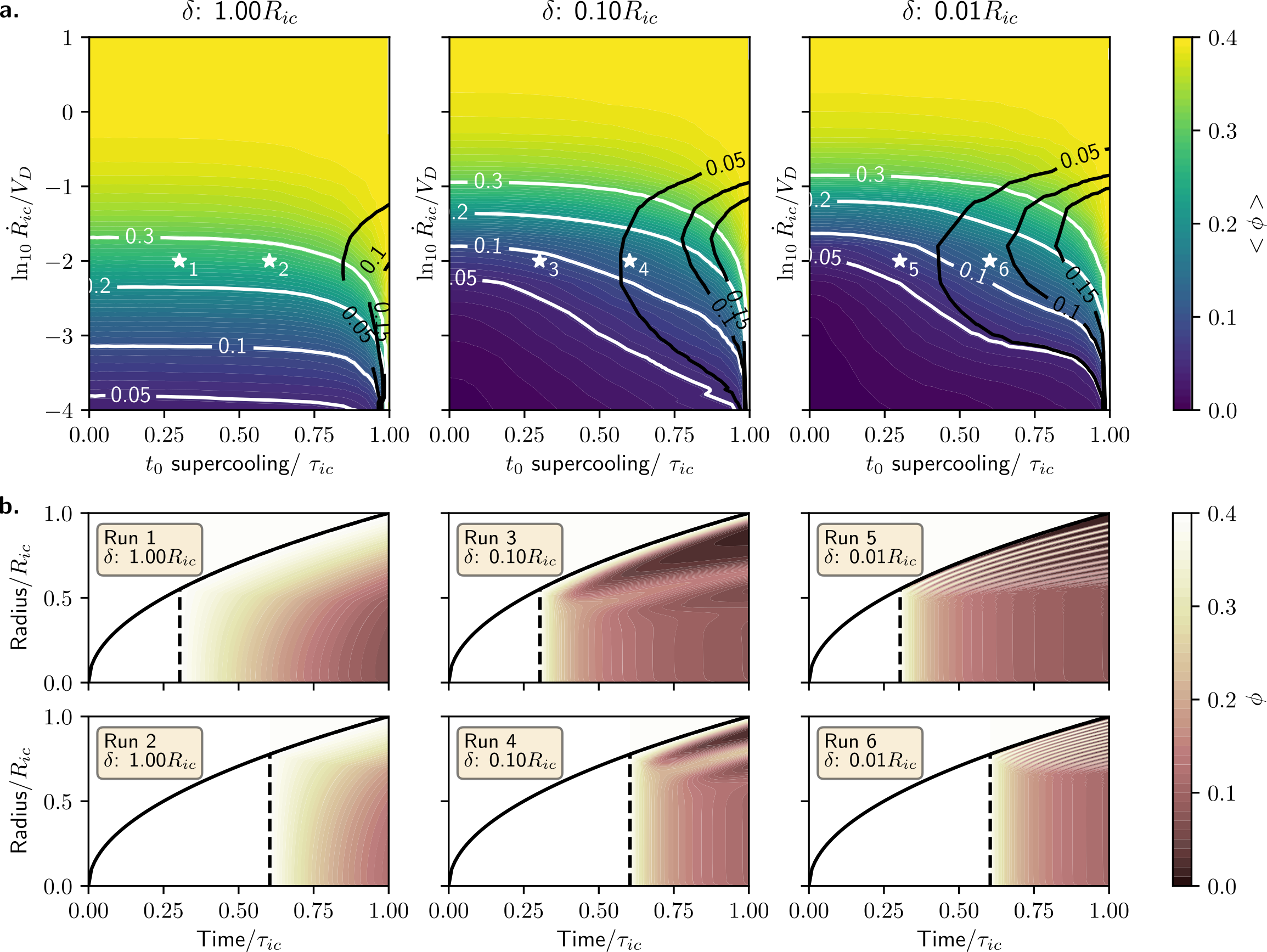}
    \caption{[a] Final average porosity for compaction lengths of $R_\text{ic}$, $0.1\,R_\text{ic}$ and $0.01\,R_\text{ic}$, as function of the supercooling time and the growth rate at the end of the run. The white lines emphasize the isosurface of average porosity, also shown in color, while the black lines are the isovalues of $<\phi>-<\phi>_\text{ref}$, the difference between the observed final value of the run  $<\phi>$ and the final value without supercooling $<\phi>_\text{ref}$. [b] Time evolution of the porosity for the 6 different runs highlighted in the panel [a] with the white star symbol. Each column correspond to the same set-up of compaction length and final growth rate, and a start time for supercooling of 0.3$\tau_\text{ic}$ and 0.6$\tau_\text{ic}$}
    \label{fig:supercooling}
    
\end{figure}

The last scenario we explore is when the growth of the inner core starts after a supercooled event. {As shown on Figure {\ref{fig:growth}}, the nucleation starts after a given time delay, with an initial fast growth followed by  a normal $\sqrt{t}$ growth. The initial fast growth is similar to an instantaneous growth, where almost no compaction occurs.} To estimate the effect of different delays in the nucleation, we calculate the average porosity at the end of the runs for different starting times for the crystallization. A supercooling time $t_0$ of 0 corresponds to a normal growth, while a starting time close to $\tau_\text{ic}$ corresponds to a nucleation delayed such that the crystallization of the inner core happened very recently.  
We calculate the final average porosity of the inner core for three different compaction length. The results are presented {in} Figure \ref{fig:supercooling} {a}. The black lines quantify the excess porosity observed in the case where a delay in nucleation is introduced compared to a classical scenario without supercooling. 
%highlights the positions where the final porosity is different from the expected final porosity without supercooling. 
This provides us with a range of parameters where the supercooling is expected to change the final porosity compared to the traditional growth history without delayed nucleation.

The effect of supercooling on the average porosity is the largest for intermediate growth rates, as it is non-visible when the growth rate is large enough to prevent compaction. For the three cases studied here, this leads to a maximum effect of supercooling for final growth rates about $\dot R_\text{ic}=10^{-2}\, V_D$, two orders of magnitude lower than the Darcy velocity. {As highlighted on Figure {\ref{fig:supercooling}} by the black solid lines,  a}
change of 5\% of the average porosity can be obtained due to an event of supercooling at this growth rate if the delayed nucleation starts after respectively $0.8\, \tau_\text{ic}$, $0.6\,\tau_\text{ic}$ or $0.5\, \tau_\text{ic}$ for compaction lengths of $R_\text{ic}$, $0.1\,R_\text{ic}$ and $0.01\,R_\text{ic}$. {Since the average porosity increases with compaction length, the amplitudes of the changes imposed by the supercooling are larger for the case $0.01\,R_\text{ic}$: in this case, a supercooling of $0.5\, \tau_\text{ic}$ corresponds roughly to a 100\% increase in porosity compared to the canonical growth.}
{Figure {\ref{fig:supercooling}}[b] presents the evolution of the porosity structure for the cases of supercooling highlighted by the white stars on  panel a. The delayed nucleation creates two different regions: an innermost volume which size corresponds to the radius of the initial fast growth where  porosity is constant and decreases with time, and an uppermost volume where solitary waves can propagate upwards.}

\section{Discussion}

In this work, we model the dynamics of two-phase flows for the compaction of a mushy inner core, exploring a large range of the two controlling parameters, the compaction length $\delta = \sqrt{k_0\mu_m/\mu_f}$ and the Darcy velocity $V_D= | \Delta \rho | g_0 k_0/\mu_f $. We also vary the growth rate with time to estimate the effect of history on the final structure. We determined scaling laws for the volume of trapped liquid and the thickness of the uppermost mushy layer for canonical growth scenarios and delayed crystallization. The physical parameters relevant for the Earth's inner core are largely unknown, and we present here how to constraints them by comparing the results of our calculations and the seismic observations, as well as the implications for the supercooling hypothesis.

% \remove{We perform all the calculations in this work with the permeability exponent set at $n=3$ to be consistent with the previous work by \protect{\citeA{Sumita1996AGrowth}}. The value of $n$ in the two-phase flow dynamics is debated, as direct measurements of permeability of partially molten rocks are impossible to obtain at high pressures and no study has been done for the specific case of iron at high pressure and high temperature. Studies rely on simple geometric models that assume pore space is connected, isotropic, and homogeneous, and that pore shape changes self-similarly. These models suggest $k \sim \phi^n$, where n is 2 or 3 for tubular or crack-like conduits \protect{\cite{gueguen1994introduction}}. Modifying the exponent to $n=2$, the results presented here are only slightly different. The porosity value is about half the values we proposed here, and the uppermost layer is of similar thickness. For magma dynamics,  $k \sim \phi^n (1-\phi)^m$ has also been proposed \protect{\cite{keller2013numerical}}, but would not drastically change our results at low porosity. }
%\note{We decided to remove this paragraph, as we consider it to be more on the technical side. We added it to the Supporting Info, combined with the figures related to the different exponent $n=2$.}

The physical parameters for metal alloy at the pressure and temperature of the Earth's inner core are not well constrained. We show here that, by considering the seismic observations of a thin uppermost mushy layer at the top of the inner core ($<5\,km$) and the small volume fraction of liquid in the bulk of the inner core ($<10$\%),   we can roughly estimate their values. Both of the compaction length and the Darcy velocity depend on the material constituting the solid matrix and the liquid in the pores, and are only marginally dependent on the size of the system -- through $g_0$. The uncertainties are mostly coming from the solid inner core viscosity and the permeability. The viscosity of the {iron alloy at inner core pressure and temperature} is difficult to constrain {\cite{bergman2018grain}}, with estimations usually spanning $10^{16}$--$10^{22}$\,Pa.s \cite{lasbleis2015building}. {Recently, values as low as $10^{1}$\,Pa.s have been proposed } \cite{belonoshko2019low}{, but these are incompatible with the shear modulus observations { \cite{xian2019viscoelasticity}}}.
The permeability coefficient of an iron mixture at inner core conditions is unknown. \citeA{Sumita1996AGrowth} proposed values as low as $k_0\sim10^{-18}$\,m$^2$, interpreted \add{based on the empirical estimation of the tortuosity }as typical crystal size of $a\sim \sqrt{1000 k_0}\sim 10^{-8}$\,m.

In our calculation, the thickness of the uppermost mushy layer $\delta _\text{ul}$ is affected only by the recent growth history of the inner core, making it independent on the growth scenario. It does not depend on the value of the permeability and scales as 
\begin{equation}
    \delta _\text{ul} \sim \delta \left ( \frac{\dot{R}_\text{ic}}{V_D}\right ) = \sqrt{\frac{\dot{R}_\text{ic}\mu_m}{|\Delta \rho |g_0}}.
\end{equation}
For $r\sim \sqrt{t}$, the final growth rate is a function of the considered age of the inner core. For an age of the inner core from 1.5\,Gyears to 0.2\,Gyears, the final growth rate is  $10^{-10}$--$1.3\cdot 10^{-11}$\,m/s. With $\dot{R}_\text{ic}=10^{-11}$\,m/s, $|\Delta \rho |=600$\,kg.m$^{-3}$ and $g_0=4.4$\,m.s$^{-2}$, we have 
\begin{equation}
    \delta _\text{ul} \sim 615\text{\,m} \ \left ( \frac{\mu_m}{10^{20}\text{\,Pa.s}}\right )^{1/2}\left ( \frac{\dot R_\text{ic}}{10^{-11}\text{\,m.s}^{-1}}\right )^{1/2}.
\end{equation}
An inner core viscosity larger than $10^{22}$\,Pa.s would promote a mushy layer at the inner core boundary larger than 6km, thick enough to be visible by seismic observation, and, as such, considered as too large to fit the current observations of the inner core.

From our calculations, the choice of scaling laws for the average porosity depends on the compaction length and the permeability. We consider a viscosity of the liquid of $10^{-2}$\,Pa.s \cite{poirier1988transport, dobson2000situ}. The Darcy velocity is 
\begin{equation}
    V_D = \frac{| \Delta \rho | g_0 k_0}{\mu_f}.   % = 1.6\cdot10^{-10}\text{\,m.s}^{-1} \left ( \frac{k_0}{10^{-15}\text{\,m}^2}\right ), 
\end{equation}
{From our calculations and as visible in Figure {\ref{fig:compaction}}, }we
find that porosity reaches values less than 10\% for a final growth rate lower than 0.03$V_D$, which corresponds to a lower bound for the permeability coefficient of $1.2\cdot 10^{-15}$\,m$^2$ with the same values as previously. 

With this limit value, the compaction length is 
\begin{equation}
    \delta = \sqrt{\frac{k_0\mu_m}{\mu_f}} = 3.1\text{\,km}\left ( \frac{k_0}{10^{-15}\text{\,m}^2}\right )^{1/2}\left ( \frac{\mu_m}{10^{20}\text{\,Pa.s}}\right )^{1/2}.
\end{equation}
To reach a compaction length larger than the inner core size, the viscosity or the permeability coefficient have to be be 6 orders of magnitude larger than the ones considered here. Assuming a compaction length smaller than the typical size of the core, the average porosity is obtained from the scaling laws as
\begin{equation}
    <\phi> = 3.3 \frac{\dot R_\text{ic}}{V_D} = 3.3 \frac{\dot R_\text{ic}\mu_f}{| \Delta \rho | g_0 k_0} = 0.13 \left ( \frac{k_0}{10^{-15}\text{\,m}^2}\right )^{-1}
    \left ( \frac{\dot R_\text{ic}}{10^{-11}\text{\,m.s}^{-1}}\right ).
\end{equation}

From seismic observations, the porosity in the inner core may be up to 10\% of liquid \cite{singh2000presence}. This constraint reduces the possible value for the permeability coefficient to $1.3$ -- $4\cdot 10^{-15}$\,m$^2$. {T}his would represent micrometric crystals in the inner core, several orders of magnitude larger than the crystal size proposed by {\citeA{Sumita1996AGrowth}}, but smaller than estimates from grain growth \protect{\cite{bergman1998estimates, deguen2007existence, yamazaki2017grain}}. The difference may lay in the definition of the permeability coefficient and \change{its}{the empirical} relationship\change{ to }{ between the tortuosity coefficient and the }grain and aggregate typical sizes. With $k_0=4\cdot 10^{-15}$\,m$^2$ and $\mu_m=10^{20}$\,Pa.s, the predicted values for the inner core of the different discussed parameters are $\delta=6.3$\,km, $V_D=1.1\cdot 10^{-9}$m.s$^{-2}$,  $\delta_\text{ul}=615$\,m, $<\phi>=3\%$.

Heterogeneous nucleation of the inner core requires the existence of an initial nucleus. \citeA{Huguet2018EarthsParadox} have shown that the supercooling may lead to a delayed crystallization, which timing is difficult to estimate. We find here that a delayed crystallization may lead to an increase in average porosity in the inner core because of the accumulation of liquid during the first step of growth.  Considering $\delta=0.01\,R_\text{ic}$ and $\dot{R}_\text{ic}\sim10^{-2}\,V_D$, we find that to trap at least an additional 5\% of liquid in the inner core -- and reach above 10\% of liquid volume-- the delay in nucleation has to be at least half the presumed age of the inner core. For a nucleation starting before that time, we expect no noticeable difference in porosity. This provides an upper bound for a reasonable timing for the start of crystallization. The initial fast growth step cannot be larger than 70\% of the radius of the inner core,  which corresponds to a supercooling of 100\,K \cite{Huguet2018EarthsParadox} and a delay in nucleation compared to the time where the first crystal may have  nucleated of about half the presumed age of the inner core in the canonical model.

The internal porosity structure may also be compared to seismic observations. For compaction length smaller than the typical size of the core, the final structure presents layers of larger porosity that are propagating upwards as solitary waves. A  central part develops, which size is typically increasing with {the} growth rate. For delayed nucleation, the central part is directly the volume of the inner core just after the fast growth step and is overlaid by a concentrated solitary waves zone (see Figure \ref{fig:supercooling} {b}). The observation of an inner-most inner core of 300 to 600\,km-radius \cite{ishii2002innermost,wang2015equatorial} may then be an evidence for an event of delayed nucleation before 0.4\,$\tau_\text{ic}$, which would be unnoticeable in term of average porosity based on previous arguments but may create a different texture in and out of the central region.

\section{Conclusion}

We develop here a 1D model to follow the structure and average porosity of the compaction of the inner core, exploring the various growth {histories} that have been proposed in past studies. We show that the amount of melt trapped in the bulk of the inner core is a record of the growth history of the inner core, and by thus a record of the thermal and compositional history of the Earth.

The possibility of a delayed nucleation of the inner core has been introduced by \cite{Huguet2018EarthsParadox}, based on the difficulty for homogeneous nucleation to occur in the well-mixed outer core. So far, no evidence for such  late nucleation {has} been found in the geomagnetic history of the Earth. Here, we propose that the porosity structure of the inner core records some of the key features of the growth history of the inner core. The uppermost layer of the inner core is sensible to the recent growth history only, while the bulk of the inner core records the past. We find that the existence of a delayed nucleation would increase the volume fraction of fluid trapped inside the inner core, and that it would likely be visible to seismic observations if the supercooling is larger than 100K, corresponding to a first growth step up to 70\% of the inner core radius today. 

A more detailed porosity structure of the inner core from seismic observations would help us constrain both the physical parameters of the inner core, through the thickness of the uppermost mushy layer, and the delayed nucleation if it existed. 
However, we note that our model lacks two main physical ingredients that should be investigated in the future: the remelting and freezing due to thermal and compositional effects and the {interstitial} convection \cite{Huguet2016StructureCore}. Both of these processes are likely to lead to a smaller melt fraction in depth. 

Several authors have proposed other exotic growth scenarios, from an inner core crystallized during accretion \cite{arkani2017formation} to several episodes of inner core growth \cite{andrault2016deep}. These scenarios would decrease the average growth rate and provide a larger time for the liquid to be expelled from the interior, leading to a lower porosity inside the inner core.

\acknowledgments
The numerical code used in this work is written in Python 3.6 and made available online on https://github.com/MarineLasbleis/mushdynamics under a GnuGPLV3 licence, as well as the installation requirements and procedures. All the figures and results in this publication use version 1.0 of the code.
This project started as MK was visiting ML at the Earth-Life Science Institute. This work acknowledges the financial support from the JSPS KAKENHI Grant Number JP18K13632,  from R\'egion Pays de la Loire, project GeoPlaNet (convention No. 2016-10982) and from  the  European Union's Horizon 2020 research and innovation program under the Marie Sk\l{}odowska-Curie Grant Agreement No. 795289.

%% ------------------------------------------------------------------------ %%
%% References and Citations

%%%%%%%%%%%%%%%%%%%%%%%%%%%%%%%%%%%%%%%%%%%%%%%
%
% \bibliography{<name of your .bib file>} don't specify the file extension
%
% don't specify bibliographystyle
%%%%%%%%%%%%%%%%%%%%%%%%%%%%%%%%%%%%%%%%%%%%%%%

\bibliography{bibliography.bib}

%Reference citation instructions and examples:
%
% Please use ONLY \cite and \citeA for reference citations.
% \cite for parenthetical references
% ...as shown in recent studies (Simpson et al., 2019)
% \citeA for in-text citations
% ...Simpson et al. (2019) have shown...
%Geophysical
%
%...as shown by \citeA{jskilby}.
%...as shown by \citeA{lewin76}, \citeA{carson86}, \citeA{bartoldy02}, and \citeA{rinaldi03}.
%...has been shown \cite{jskilbye}.
%...has been shown \cite{lewin76,carson86,bartoldy02,rinaldi03}.
%... \cite <i.e.>[]{lewin76,carson86,bartoldy02,rinaldi03}.
%...has been shown by \cite <e.g.,>[and others]{lewin76}.
%
% apacite uses < > for prenotes and [ ] for postnotes
% DO NOT use other cite commands (e.g., \citeA, \cite, \citeyear, \nocite, \citealp, etc.).
%

\end{document}